# PRELIMINARY CONSIDERATIONS ABOUT THE INJECTORS OF THE HE-LHC

R. Garoby, CERN, Geneva, Switzerland

*Abstract*

A hadron collider operating at an energy much larger than the LHC ("HE-LHC") would be a logical successor to the LHC itself, especially if its cost can be minimized by reusing a significant part of the CERN infrastructure like the existing tunnels and/or accelerators. The injector complex must however be extended to reach a beam energy of ~1.2 TeV and. in view of the time span of the HE-LHC project, the replacement of ageing accelerators can also be necessary. The main possible options are outlined together with their specificities.

## INTRODUCTION

Beyond the need to satisfy the requirements of the HE-LHC, the options for the injector complex have to take into account the peculiarities of the existing accelerators and their other present or potential uses for physics. Moreover, choices have to result from an overall optimization of the whole HE-LHC project, taking into account the cost of dismantling and operation, as well as the opportunities offered by decommissioned installations like HERA, Tevatron and the LHC itself.

## HE-LHC REQUIREMENTS

Preliminary considerations for a higher-energy LHC ("HE-LHC") [1] have lead to the figures listed in Table 1 for the beam characteristics at injection in the future collider.

Table 1: Beam characteristics in the HE-LHC

| | |
|---|---|
| Injection energy | ~1.2 TeV |
| Protons/bunch | $1.3 \times 10^{11}$ |
| Time interval between bunches | 50 ns |
| Transverse emittances (H/V) | 3.75 / 1.84 |
| | or 2.59 / 2.59 μrad |
| Longitudinal emittance | < 4 eVs |

The maximum energy of the SPS being of 450 GeV, a new accelerator is required in the injector chain to reach 1.2 TeV. All other requirements can be met by the CERN injector complex especially after its upgrade for the High Luminosity LHC ("HL-LHC") [2]. Considering however that the HE-LHC will start being operational in approximately 20 years, the present accelerators may represent a reliability concern if they have not been replaced.

## PLANS FOR HL-LHC

Until the beginning of 2010, studies were taking place in view of building new accelerators to replace the PSB and PS, boosting performance and reliability of the first part of the injection chain, while simplifying the upgrade of the SPS by injecting at a much higher energy (50 instead of 26 GeV) [3, 4, 5]. This solution was however discarded after the 2010 LHC Performance Workshop in Chamonix, and the decision was taken to rather consolidate and upgrade the PSB and PS [6]. As a result, the SPS will keep operating with maximum injection energy of 26 GeV and the accelerators in the injector complex will date from 55 and 70 years when HE-LHC will start, except Linac4 which will be only ~14 years old.

## OPTIONS FOR THE INJECTORS

Three main options are being considered for delivering beam at 1.2 TeV to the HE-LHC:

- New synchrotron using superconducting magnets in the SPS tunnel;
- New synchrotron using low field superconducting magnets in the LHC tunnel;
- LHC as a pre-accelerator.

### New synchrotron in the SPS tunnel

Assuming the same filling factor than in the SPS, the maximum $B$ field in the dipoles have to be of ~5.5 T. To provide the same proton flux at ejection (same filling time of the collider and same flux for fixed target physics), the $dB/dt$ has to be of ~1.8 T/s. Such characteristics are close to the ones of the SIS300 dipoles for the FAIR project [7].

Injection in this new synchrotron has to be at ~100 GeV. If the 1.2 TeV accelerator can co-exist with the SPS in the existing tunnel, the possibility to use the SPS as injector could be considered. This is certainly difficult, but it cannot be rejected before some serious study. It is however more likely that a new machine will be preferable, either with a small footprint in the SPS tunnel, or in a new and shorter tunnel. The needs of this new accelerator will impact upon its injectors and significant work has to be invested for studying the options, especially if the requirements of other physics users have to be taken into account.

Another important consequence results from the need to rebuild the transfer lines TI8 and TI2 to the LHC (5.6 km in total). The possibility to use HERA or Tevatron magnets for that purpose is commented upon during this workshop [8, 9].

*New synchrotron in the LHC tunnel*

If acceleration above 450 GeV can be done by a machine located in the LHC tunnel, the transfer lines do not need to be modified and the SPS as well as the lower energy accelerators could in principle stay untouched. This solution requires however:

- The addition of a new dual beam synchrotron within the tight space left available by the collider;
- Finding a means to pass through or to by-pass the detectors.

This possibility has already been subject to a preliminary study [9] and it is addressed during this workshop [10]. The dipoles of the proposed new accelerator, called LER (Low Energy Ring) are based on the technology envisaged for the VLHC. In one option, two 1 km by-pass tunnels have to be built for the LER beam-pipe to avoid passing through the large detectors in IP1 and IP5. Beam transfer from LER to HE-LHC is tentatively located in IP7. In the second option, no by-passes are necessary because the LER beam is deflected to pass every turn through the centre of the detectors in IP1 and IP5. It relies on fast deflectors which would also solve the question of injection in the HE-LHC, although making it take place in a very fragile part of the machine, namely in the centre of the detectors.

*LHC as a pre-accelerator for the HE-LHC*

The cost of the magnets for the HE-LHC is likely to be very large, even compared to the cost of building a new and longer tunnel. An economical optimization will therefore be necessary, taking into account the variation of the magnets cost as a function of the maximum bending field and the cost of constructing a new and longer tunnel. It should be added in the analysis that, if the HE-LHC is located in a new tunnel, the LHC could remain in place and be used as injector, which would also remove the need and cost of its dismantling.

This option can only be considered after the R § D on the HE-LHC magnets will have sufficiently progressed

## CONCLUSION

The HE-LHC raises two kinds of challenges for its injectors. The first one concerns beam energy, which has to be at ~1.2 TeV. It can be addressed with a new 1.2 TeV synchrotron either in the LHC tunnel or in the SPS tunnel. An alternative would be to use the LHC itself, if it is economically interesting to locate the HE-LHC in a new and longer tunnel. The second one results from the time period which covers ~2030-2050. This is likely to require the replacement of part or all of the existing synchrotrons e.g. to improve reliability, reduce the cost of maintenance (manpower and material) and decrease the environmental impact. The specifications of these new accelerators could also be influenced by the needs of other physics users (e.g. for neutrinos and/or nuclear physics).

In any case, as soon as the HE-LHC will become an attractive option for the future of particle physics, it would make great sense to start preparing the injector complex and let the HL-LHC and the other users benefit from new/ renovated accelerators.